\documentclass[twocolumn,amsmath,amssymb,aps,pra,floatfix,nofootinbib,superscriptaddress]{revtex4}
\usepackage{graphicx,graphics}
\usepackage{bm}
\usepackage{longtable}

\def\be{\begin{equation}}
\def\ee{\end{equation}}
\def\ba{\begin{eqnarray}}
\def\ea{\end{eqnarray}}
\def\la{\langle}
\def\ra{\rangle}

\def\h{\hskip 1cm}

\newcommand{\ket}[1]{|{#1}\rangle}
\newcommand{\bra}[1]{\langle{#1}|}

\DeclareMathOperator{\sign}{sign}
\DeclareMathOperator{\Tr}{Tr}
\DeclareMathOperator{\airy}{Ai}

\begin{document}
\title{Initializing an unmodulated spin chain to operate as a high quality quantum data-bus}

\author{Abolfazl Bayat}
\affiliation{Department of Physics and Astronomy, University College
London, Gower St., London WC1E 6BT, United Kingdom}

\author{Leonardo Banchi}
\affiliation{Dipartimento di Fisica, Universit\`a di Firenze,
Via G. Sansone 1, I-50019 Sesto Fiorentino (FI), Italy}
\affiliation{INFN Sezione di Firenze, via G.Sansone 1, I-50019 Sesto
Fiorentino (FI), Italy}

\author{Sougato Bose}
\affiliation{Department of Physics and Astronomy, University College
London, Gower St., London WC1E 6BT, United Kingdom}

\author{Paola Verrucchi}
\affiliation{ISC - Consiglio Nazionale delle
Ricerche, UoS via G.Sansone 1, I-50019 Sesto Fiorentino (FI), Italy}
\affiliation{Dipartimento di Fisica, Universit\`a di Firenze,
Via G. Sansone 1, I-50019 Sesto Fiorentino (FI), Italy}
\affiliation{INFN Sezione di Firenze, via G.Sansone 1, I-50019 Sesto
Fiorentino (FI), Italy}

\begin{abstract}
We study the quality of state and entanglement
transmission through quantum channels described by spin chains
varying both the system parameters and the initial state of the channel.
We consider a vast class of one-dimensional many-body models
which contains some of the most relevant experimental realizations of
quantum data-buses.
In particular, we consider spin-$1/2$ XY and XXZ model with open boundary
conditions.
Our results show a significant difference between free-fermionic (non-interacting) systems (XY)
and interacting ones (XXZ), where in the former case initialization can
be exploited for improving the entanglement distribution, while in the latter
case it also determines the quality of state transmission.
In fact, we find that
in non interacting systems the exchange with fermions in the initial state of the chain always has a destructive effect, and we prove that it can be completely
removed in the isotropic XX model by initializing the chain in a ferromagnetic
state. On the other hand,
in interacting systems constructive effects can arise by scattering between hopping fermions and a proper initialization procedure.
Remarkably our results are the first example in which state and
entanglement transmission show maxima at different points as the interactions and initializations of spin chain channels are varied.
\end{abstract}

\date{\today}
 \maketitle

\section{INTRODUCTION} \label{Introduction}

Quantum communication between different registers/processors is a vital
task in fast developing quantum technology. Using mobile particles, such
as photons or moving electrons, for carrying information is one option
which either faces the complexity of interface equipments for different
physical objects (e.g. photons and electrons) or needs a fine control
over the bus which is still challenging \cite{taylor,petrosyan}. An
alternative is to let the information {\it flow} through a quantum
channel, physically realized by a chain of permanently coupled localized
particles, exploting the dynamical properties of the channel itself.
There are several possibilities for realizing channels that might serve
this purpose, amongst which spin-1/2 chains have revealed particularly
suitable for transferring quantum information from one point to another
\cite{bose,bose-review,bayat-XXZ}.

Depending on the specific physical realization of the overall system, it
might be easier to act on the structure of the Hamiltonian ruling the
channel dynamics or to prepare the channel in a specific initial state.
For instance, spin chains in solid state physics represent a vast
reservoir of possible quantum channels, characterized by the most
diverse Hamiltonians, though with fixed parameters
\cite{SteinerVW76,mikeska}. On the other hand,
initializing a spin chain
embedded on a solid-state matrix might be a hard task.
Quite complementary, recent progress in optical lattices are
making a real chance out of several theoretical proposals for realizing
spin chains with cold atoms
\cite{Bakr2009,Brennen1999,Mandel2003,Greiner2002,Sherson2010Bakr2010},
though with some restrictions on the structure of the effective spin
Hamiltonians actually attainable \cite{Lukin2003}. Moreover, different
initial states can be realized in an optical lattice
\cite{weitenberg-ferro,Koetsier-neel,Barmettler-singlet}, and new
cooling techniques \cite{Medley} also provide the possibility of
reaching temperatures in which the magnetic phases are not disturbed by
thermal fluctuations and so the real magnetic ground state of the system
becomes reachable.

There are two essential features that characterize quantum channels made
of interacting localized objects: their dynamics is dispersive, due to
the non trivial structure of the many-body Hamiltonian that describes
the channel, and it depends on the initial state of the channel itself.
Dispersion is always detrimental to quantum information transmission,
and designing a non dispersive channel requires a detailed engineering of the
local couplings\cite{christandl,DiFranko-Perfect}, which is
practically hard to achieve.
It is therefore relevant to understand up to what extent according to
the type, parameters and the length, a homogeneous non locally
engineered spin chain is usable for quantum communication. In
particular, and at variance with some recent works in which high quality
transmission is achieved independent of the system initialization
\cite{DiFranko-initial,Yao,BACVV2010}
we would like to see whether or not one can
improve the quality of transmission by means of a specific
initialization. Another issue which is less studied in the literature
(unless very few cases for the case of engineered chains
\cite{DiFranko-Perfect,DiFranko-initial}) is the effect of Hamiltonians
which do not conserve the number of excitations. In fact our
investigation here includes a wide class of Hamiltonians which change
the number of excitations during the time evolution.

We consider quantum channels realized by finite
spin-1/2 chains with homogeneous nearest neighbor exchange interaction of the
Heisenberg type, possibly in the presence of a uniform external magnetic
field.
As for the initial state of the channel we consider the
ferromagnetic state, with all the spins parallel
to each other, the N\`eel state, where the spins are
alternatively parallel, the state built as a series of singlets, and the
ground state.
In order to study the interplay between the properties of the
channel Hamiltonian and the structure of the initial state in determining
the quality of the transmission processes, we specifically deal
with different Hamiltonians and different initial states.
We present a comprehensive study for the transmission quality over whole
phase diagram of the XY and XXZ Hamiltonians with the above initial
states.

The structure of the paper is as follows: in Section \ref{Setup} we
introduce our scheme for quantum state transfer and entanglement
distribution in a general language. In Section \ref{sec_XY} we
study free fermionic models, i.e. XY Hamiltonian, while the Section
\ref{XXZ} is devoted to the XXZ model as an example of interacting
systems. Finally, in Section \ref{conclusion} we comment upon our
results.

\begin{figure}
\centering
    \includegraphics[width=8cm,height=5.6cm,angle=0]{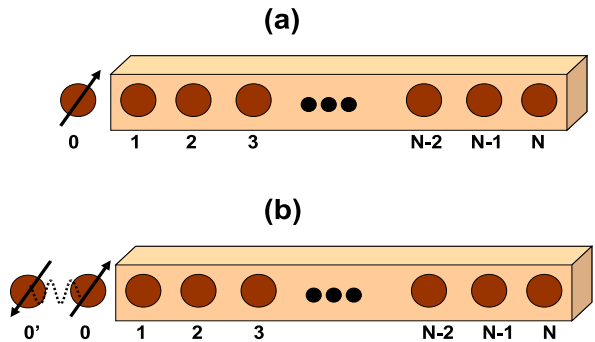}
    \caption{(Color online) Schematic picture for:
(a) State transferring; (b) Entanglement distribution.}
    \label{fig1}
\end{figure}

\section{set-up}
\label{Setup}

The quantum channel consists of $N$ spin-1/2 particles sitting at sites
$1$ to $N$ of a one dimensional lattice and interacting through the Hamiltonian
\begin{equation}
\label{Hch}
H_{\rm ch}=\sum_{l=1}^{N-1} \left(
 J_x \sigma^x_l\sigma^x_{l+1}
 + J_y \sigma^y_l\sigma^y_{l+1}
 + J_z \sigma^z_l\sigma^z_{l+1}\right)
 +h\sum_{l=1}^N \sigma_l^z, \end{equation}
where $J_\alpha$ ($\alpha=x,y,z$) are the exchange integrals, $h$ is an
external uniform magnetic field applied in the $z$ direction, and
$\sigma_l^{\alpha}$ are the Pauli operators of the spin sitting at site
$l$. We prepare the channel in some initial pure state
$|\psi_{\rm ch}\ra$, which can be either entangled or separable with
respect to single-spin states.
An extra qubit which carries the information is labelled by the site
index $0$ and initially set in some arbitrary state $\rho_{_0}(0)$.
The schematic picture of the system is shown in Fig.~\ref{fig1}(a).

%
%

At $t=0$ the interaction between the qubit and the channel is suddenly
switched on, via
\begin{equation}\label{HI}
    H_I
= J_x \sigma^x_0\sigma^x_{1}
+ J_y \sigma^y_0\sigma^y_{1}
+ J_z \sigma^z_0\sigma^z_{1}
+h\,\sigma_0^z.
\end{equation}
We use sudden switching for computational simplicity as the dynamics is
not altered by a more realistic finite switching time, provided that it
is small compared to the characteristic times set by the couplings of the
Hamiltonian $H_I$ \cite{BBVB}.

The total Hamiltonian
$H=H_{\rm ch}+H_I$ rules the dynamics of the overall system, whose state
at time $t$ reads
\begin{equation}\label{psi_t}
\rho(t) = e^{-iHt}\,\rho(0)\,e^{iHt}
\end{equation}
where $\rho(0) = \rho_{_0}(0)\otimes \ket{\psi_{\rm ch}}\bra{\psi_{\rm ch}}$ and
$\hbar=1$.
The density matrix of the qubit $N$, namely $\rho_{_{N}}(t)$,
is obtained by tracing out the qubits $0,1,...,N-1$ from $\rho(t)$.
On the other hand, one can define the superoperator $\mathcal{E}_t$ which
maps the initial density matrix of the qubit 0, i.e. $\rho_{_0}(0)$, to
the density matrix of qubit $N$ at time $t$, i.e. $\rho_{_N}(t)$, according to
\begin{equation}\label{xi}
\rho_{_N}(t)=\mathcal{E}_t[\rho_{_0}(0)].
\end{equation}
When the transmission of a generic pure quantum state $\ket{\psi_{_0}}$ is in
order, according to the scheme depicted in Fig.~\ref{fig1}(a),
we set $\rho_{_0}(0)\equiv\ket{\psi_{_0}}\bra{\psi_{_0}}$ and
quantify the quality of transmission by the fidelity
$\la \psi_{_0}|\rho_{_N}(t)|\psi_{_0}\ra$ between the initial state
$|\psi_{_0}\ra$ and
the final state $\rho_{_N}(t)$.
In fact, the quality of the channel is better assessed by the
average fidelity $\int d\psi_{_0}\la \psi_{_0}|\rho_{_N}(t)|\psi_{_0}\ra$,
	where
integration is over the surface of the Bloch sphere and $d\psi_{_0}$ the
corresponding measure.
There are essentially two mechanisms that cause average fidelity to
deteriorate during a transmission process of the type we are describing:
dispersion, which is a collective phenomenon due to the channel
being an interacting many-body system, and local rotations.
As a matter of fact, the average fidelity does not distinguish between bad
transmission, i.e. dispersion of the state all over the chain, and good
transmission with an extra rotation during the dynamics; on the other
hand, the latter
can be safely handled with an extra unitary operation on the site $N$,
thus leaving dispersion as the {\em only} destructive effect in
transmitting quantum states.

Assume a unitary operator $R$, that does not depend on the state
$|\psi_{_0}\ra$, is found such that the average fidelity for the rotated
final state $R^\dagger\rho_{_N}(t)R$ equals the maximum attainable value
through the specific channel, then one could get a quantitative estimate
of the dispersiveness of the transmission, which might be a very useful
tool for characterizing the channel suitability for state transfer
processes. Aiming at such goal, we introduce the Optimal Average
Fidelity (OAF)
\begin{equation}\label{Fav1}
F(t)=\max_{R\in U(2)} \int d\psi_{_0} \bra{\psi_{_0}}R^\dagger\mathcal{E}_t
\left[\ket{\psi_{_0}}\bra{\psi_{_0}}\right]R\ket{\psi_{_0}} ~,
\end{equation}
where maximization over $R$ guarantees that the effect of local
rotations is removed. It is of absolute relevance that, as shown in
Appendix A, the OAF can be determined explicitly in terms of the
superoperator $\mathcal{E}_t$
\begin{equation} \label{Fav2}
F(t) = \frac12 +\frac1{12}
\left(m_1+m_2+\sign(\det(\mathcal{M}^\mathcal{E}))\,m_3\right)
\end{equation}
where $m_i(t)$ are the singular values, in decreasing order, of the
matrix
\begin{equation}\label{mmat}
\mathcal{M}^\mathcal{E}_{mn}(t) =
\Tr[\sigma_m\mathcal{E}_t(\sigma_n)]~,
\end{equation}
where $\sigma_n$, $n=1,2,3$ are the Pauli matrices.
The Optimal Average Fidelity, once computed via Eq.~(\ref{Fav2}),
gives a quantitative indication about how well a channel behaves, as far as
the pure states transmission is concerned.
Notice that the above expression has the very same form of the maximal
fidelity with respect to maximally entangled states \cite{badziag}, which,
on the other hand is a completely different measure (an entanglement
measure, in fact).

When entanglement distribution comes into play, mixed state
transmission must be considered, and other strategies are necessary.
Let us prepare qubit $0$ in a maximally entangled state with an isolated
qubit $0'$, see Fig.~\ref{fig1}(b).
The dynamical evolution cause the mixed state of qubit 0 to be transmitted to
qubit $N$, thus generating entanglement between qubit $0'$ and $N$.
We can quantify the quality of such transmission by the amount of
entanglement between qubits $0'$ and $N$ at time $t$. For a generic spin
chain, when qubits $0$ and $0'$ are initially in the maximally entangled
state $|\Phi^+\ra=\left(\ket{00}+\ket{11}\right)/\sqrt2$ one can write
\begin{equation}\label{rho_0N}
\rho_{_{0'N}}(t)=(I\otimes \mathcal{E}_t)\left[|\Phi^+\ra \la \Phi^+|\right],
\end{equation}
where, $I$ represents the identity map. In the following, we use concurrence
$C(\rho)$
as an entanglement measure \cite{concurrence} for quantifying the amount of
entanglement shared between qubits $0'$ and $N$.
Notice that different choices of initial maximally entangled state
would give the same result, as the set of maximally entangled state can
be obtained from $\ket{\Phi^+}$  through a local unitary operation in
$0'$, which is isolated.

Consistently with the transfer process, our scheme implies the existence
of an arrival time when both the optimal average fidelity $F(t)$
and the concurrence $C(t)=C(\rho_{_{0'N}}(t))$ get their maximum value.
In fact,
due to the finite size of the chain, the information travels from $0$ to $N$
and viceversa multiple times, and the above quantities displays multiple peaks
during the dynamics.
Throughout this paper we concentrate on the first peak,
whose position defines the arrival time $t=t^*$,
as in a practical situation waiting for longer times is unwise due to the effect of decoherence.

\section{FREE FERMIONIC SYSTEMS: XY HAMILTONIAN } \label{sec_XY}

In this section we consider the XY model defined by Eq.~\eqref{Hch} with
\begin{equation}\label{XY}
    J_x=J\frac{1+\gamma}{2},\ \ \
    J_y=J\frac{1-\gamma}{2},\ \ \
    J_z=0,
\end{equation}
where, $J$ is the exchange coupling and $\gamma$ is the anisotropy parameter.
This model is exactly solvable, as it turns into a free fermionic system,
which makes valuable analytical results available. Despite being mapped into
a non-interacting system, the model in the infinite $N$ limit
has a rich phase diagram  featuring a
quantum phase transition \cite{sachdev}
at $h=1$ and, as far as the entanglement properties
are concerned, the divergence of the entanglement range
when approaching the curve $h^2+\gamma^2=1$, where
pairwise entanglement vanishes \cite{paola1,paola2,paola3}.
Moreover, its peculiar non equilibrium dynamics has been studied in the
framework of dynamical entanglement sharing  \cite{osterloh}, with periodic
boundary conditions assumed.

For diagonalizing this Hamiltonian one first maps spin operators $\sigma^{\pm}=(\sigma^x\pm i \sigma^y)/2$  to fermionic operators through the Jordan-Wigner
transformation
$c_l=\prod_{n=0}^{l-1} \left(-\sigma_n^z\right) \sigma_l^- $
where $\{c_l,c_{l'} \}=0$ and $\{c_l,c_{l'}^\dagger \}=\delta_{l,l'}$,
as can be easily proven.
The fermionic Hamiltonian can then be diagonalized with the procedure
described in \cite{LiebSM1961}, of which we give a short summary in
appendix B. The resulting diagonal Hamiltonian is
\begin{equation}\label{Q_diag}
H=\sum_{k=0}^N E_k d_k^\dagger d_k,
\end{equation}
where the diagonal fermionic operators are obtained via the
Bogolubov transformation
\begin{equation}\label{eta}
d_k=\sum_{l=0}^N P_{_{kl}} c_l + Q_{_{kl}} c_l^\dagger.
\end{equation}
In the here considered case of finite $N$ and open boundary conditions
the analytical expressions of the energies $E_k$ and of the matrices $P$, $Q$
for finite $\gamma$ and $h$ are complicated \cite{loginov},
but they can be determined numerically as explained in appendix B.

The channel can be conveniently characterized in terms of
its dynamical behaviour, which is investigated in the Heisenberg
representation.
The fermionic operators $c_l(t)=e^{+iH t}c_l e^{-iH t}$ read
\begin{equation}\label{ck_t}
c_l(t)=\sum_{n=0}^N U_{ln}(t)c_n+W_{ln}(t)c_n^\dagger,
\end{equation}
where,
\begin{align}\label{W-U}
U(t)=& P^Te^{-itE}\,P + Q^Te^{+itE}\, Q, \\
W(t)=& P^Te^{-itE}\,Q + Q^Te^{+itE}\, P,
\end{align}
$E$ is the diagonal energy matrix with elements $E_k$,
and the inverse transformation Eq.~\eqref{inverse-eta} together with
the identity $e^{+iH t}d_k e^{-iH t}=e^{-itE_k}d_k$ have been used.
From the dynamical evolution of $c_l(t)$ that of the last-spin operators
follow
\begin{equation}\label{sigmant}
\sigma_N^-(t)=\prod_{k=0}^{N} \left(-\sigma_k^z\right)\left[\sum_{l=0}^N U_{Nl}(t)c_l+W_{Nl}(t)c_l^\dagger\right],
\end{equation}
where conservation of the parity $\prod_{l=0}^{N} (-\sigma_l^z)$ is used.

We prepare the qubit $0$ in the density matrix  which is
parametrized by the vector $\vec n(0)$ of the Bloch sphere via
$\rho_0(0) = (I + \vec n(0) \cdot \vec\sigma)/2$. Correspondingly,
the time evolution of the state of qubit $N$,
$\rho_N(t) = (I + \vec n'(t) \cdot \vec\sigma)/2$, is parametrized
by the vector
$n'_{\alpha}(t) = \bra{\Psi(0)}\sigma^{\alpha}_{N}(t)\ket{\Psi(0)}$,
which is determined from \eqref{sigmant}.
Assuming that the channel is initialized in some state with constant parity $p$, i.e. $\prod_{k=1}^N \left(-\sigma_k^z\right) |\psi_{\rm ch}\ra=(-1)^p |\psi_{\rm ch}\ra$, our quantum channel \eqref{xi} is described by the following affine
transformation which maps vector $\vec n(0)$ of the Bloch sphere of
qubit 0
to vector $\vec n'(t)$ of the Bloch sphere of qubit $N$:
\begin{eqnarray}\label{nx-ny-nz}
\begin{pmatrix} n'_x \\ n'_y \end{pmatrix} &=&
S_{\phi_w-\phi_u} \begin{pmatrix} u-w&0 \\ 0 & u+w \end{pmatrix}
S_{\phi_w+\phi_u}^T
\begin{pmatrix} n_x \\ n_y \end{pmatrix} \cr \cr
n'_z \phantom{\Big)} &=&  n_z\big(u^2-w^2\big)+u^2+w^2 +2A(t)-1
\end{eqnarray}
where $u(t) = |U_{N0}(t)|$, $w(t) = |W_{N0}(t)|$,
$\phi_u(t) = \arg[U_{N0}(t)] + (p+1)\pi$,
$\phi_w(t) = -\arg[W_{N0}(t)] + p\pi$,  the rotation matrix is
\begin{equation} \label{rotation_matrix_R}
  S_{2 \phi} = \begin{pmatrix} \sin\phi&\cos\phi \\
-\cos\phi & \sin\phi \end{pmatrix},
\end{equation}
and $A(t)$ is the real function
\begin{align}\label{A_t}
    A(t) = \sum_{j,l=1}^N & U_{Nj}(t)^*U_{Nl}(t)\la \psi_{\rm ch}|c_j^\dagger c_l |\psi_{\rm ch}\ra +
    \\\nonumber &
    W_{Nj}(t)^*W_{Nl}(t)\la \psi_{\rm ch}|c_j c_l^\dagger |\psi_{\rm ch}\ra +
    \\\nonumber &
    U_{Nj}(t)^*W_{Nl}(t)\la \psi_{\rm ch}|c_j^\dagger c_l^\dagger |\psi_{\rm ch}\ra +
    \\\nonumber &
    W_{Nj}(t)^*U_{Nl}(t)\la \psi_{\rm ch}|c_j c_l |\psi_{\rm ch}\ra.
\end{align}
Eqs.~\eqref{nx-ny-nz} show that the evolution in the $xy$-plane
involves
the two rotations $S_{\phi_w-\phi_u}$, $S_{\phi_w+\phi_u}^T$  as well as
the shrinking towards the center of the
Bloch sphere embodied in $u(t)-w(t)$ and $u(t)+w(t)$. Notice that
none of the quantities involved depends on the initial state
$\ket{\psi_{\rm ch}}$,
thus relating the dynamics in the $xy$-plane only with the phase-diagram
$\gamma$-$h$ of the model.
In fact, the only dependence on the initial state is in the quantity $A(t)$
which uniquely affects the shift in the $z$ direction, and represents the
interference of $\ket{\psi_0}$ with $\ket{\psi_{\rm ch}}$ during the evolution.
Notice that in the XX case, where $W_{ij}(t)=0$ (see appendix C), the
dynamics \eqref{nx-ny-nz} realizes a generalized amplitude damping channel
\cite{NielsenC2000}.

The OAF and the Concurrence can be calculated once the superoperator
$\mathcal{E}_t$ is known.
The elements of the superoperator $\mathcal{E}_t$ can be read from
\begin{equation}
\la i|\rho_{_N}(t)|j\ra=\sum_{k,l=0,1}
\la i| \mathcal{E}_t(|k\ra \la l|) |j\ra \la k|\rho_{_0}(0)|l\ra.
\end{equation}
From the elements $\la i| \mathcal{E}(|k\ra \la l|) |j\ra$
we can construct the Choi matrix $\mathcal{C}_\mathcal{E}(t)$
\begin{equation}\label{choi}
\bra{ki}\mathcal{C}_\mathcal{E}(t)\ket{lj} = \la i|\mathcal{E}_t(|k\ra \la l|)|j\ra
\end{equation}
which completely characterizes \cite{choi} both the superoperator
$\mathcal{E}_t$ and the state \eqref{rho_0N}.
By explicit calculation we have
\begin{widetext}
\begin{equation}\label{choixi}
\mathcal{C}_\mathcal{E}(t)  =\left(
\begin{array}{cccc}
u(t)^2 + A(t) &0&0& u(t)e^{i\phi_u(t)} \\
0& 1-u(t)^2-A(t) & w(t)e^{i\phi_w(t)} & 0 \\
0&w(t)e^{-i\phi_w(t)} & w(t)^2+A(t)&0 \\
u(t)e^{-i\phi_u(t)} & 0 &0 & 1-w(t)^2-A(t)
\end{array}
\right).
\end{equation}
\end{widetext}

From the Choi matrix,  using the procedure described
in appendix A, we can compute the OAF: the singular values of $\mathcal{M}_\mathcal{E}$(t) are
$2|u(t)+w(t)|$, $2|u(t)-w(t)|$, and $2|u(t)^2-w(t)^2|$, from which
\begin{equation}\label{Fav3}
    F^{\rm XY}(t)=\frac12+\frac16\left|u(t)^2-w(t)^2\right|
    +\frac13\max\{u(t), w(t)\}.
\end{equation}
This is a remarkable result as it is fully {\em independent} of the
initial state of the channel (i.e. the parameter $A(t)$) and depends only
on the Hamiltonian parameters. Moreover,
the rotation that maximizes the average fidelity is found to be
\begin{equation}\label{rotationxy}
R = \begin{cases} e^{-i\frac{\phi_u}{2}\sigma^z} & \text{ for } u > w \\
e^{i\frac{\pi}{2}\sigma^x} e^{i\frac{\phi_w}{2}\sigma^z} & \text{ for } u < w
\end{cases}.
\end{equation}
In the XX case the effect of the magnetic field on the dynamics
is only in the phase $\phi_u$, as shown in appendix C, and therefore one
can always choose the magnetic field such that at the arrival time
$t^*$, i.e. when fidelity peaks, $\phi_u(t^*)=0$. On the other hand,
in the XY case the dynamical quantities depends on $h$ in a complicated
way, making the explicit rotation $R$, given in Eq. \eqref{rotationxy},
necessary.

\begin{figure}\begin{center}
\includegraphics[width=0.4\textwidth]{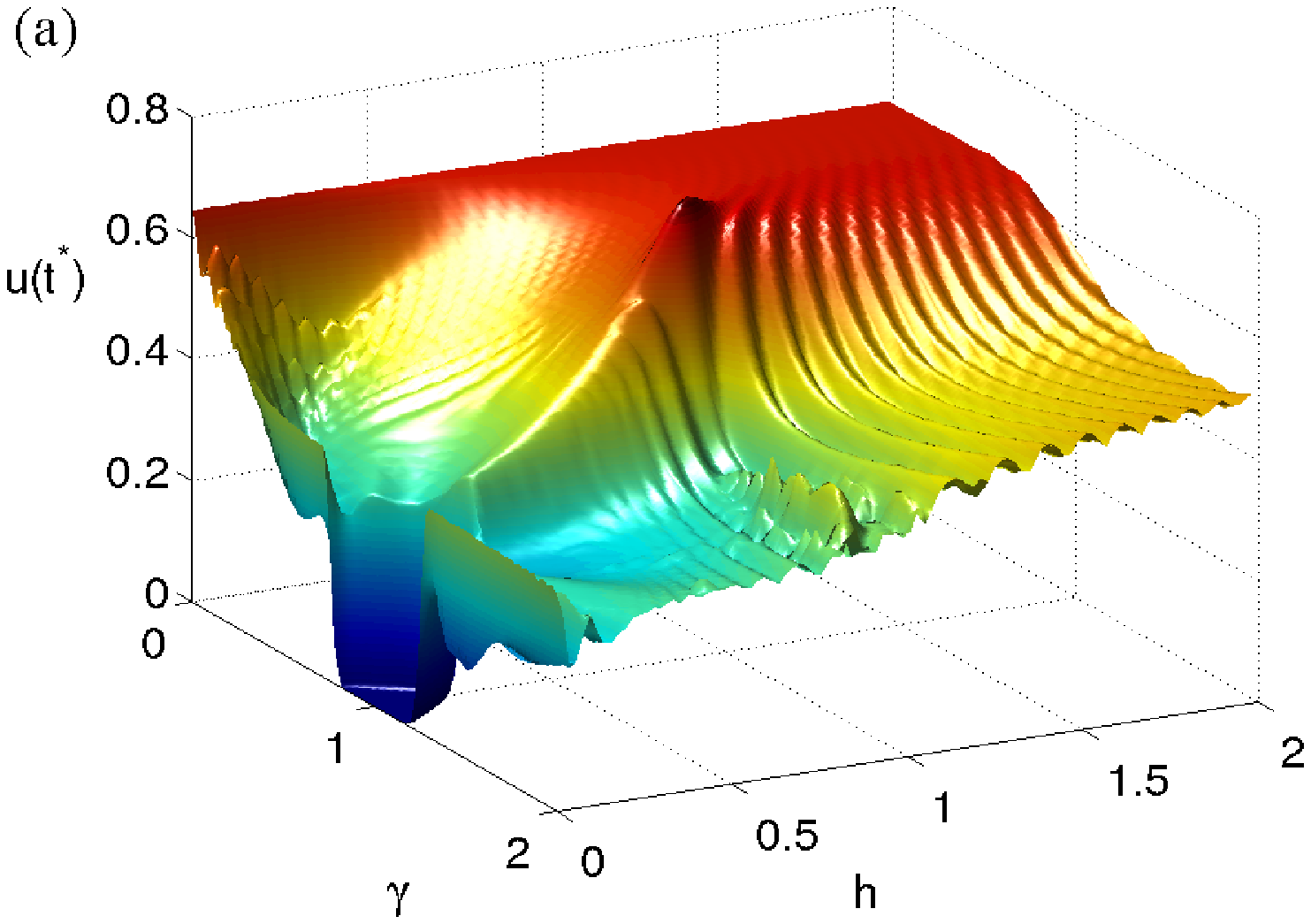}
\includegraphics[width=0.4\textwidth]{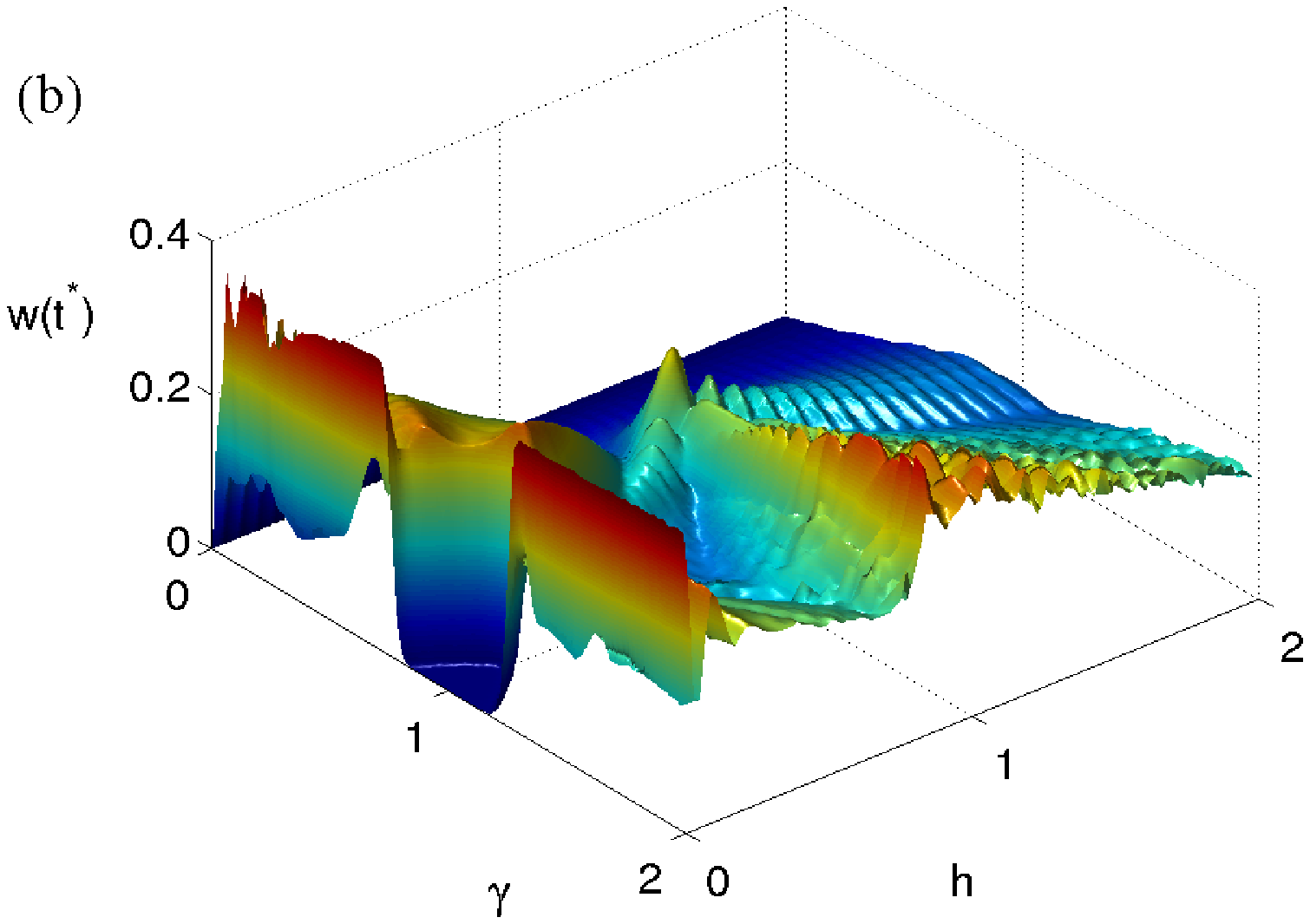}
\caption{(Color online)  (a) $u(t^*)$ vs. $\gamma$ and $h$;
the peak is located at
$\gamma=0.7$ and $h=1$.
(b) $w(t^*)$ vs. $\gamma$ and $h$. Both figures are for $N=50$.}
\label{fig2}
\end{center}\end{figure}

From Eq.~\eqref{Fav3} we see that the larger
the difference between $u(t^*)$ and $w(t^*)$ the larger the OAF.
In particular, as seen in Fig.~\ref{fig2},
we find that, whenever the fidelity is large, we see $u(t^*)\gg w(t^*)$,
so the qualitative behavior of the OAF is the same of $u(t^*)$.
Moreover, in the region $0\leq\gamma\leq 1$ and $1 \leq
h\leq 2$,  $u(t^*)$ is large, taking
its maximum value for $\gamma=0$ (XX case) and for $\gamma=0.7$ and
$h=1$. This means that in the $\gamma$-$h$ phase-diagram the line $\gamma=0$
and the point $(0.7, 1)$ set the best possible Hamiltonian parameters,
corresponding to the least dispersive channel.

\begin{figure}\begin{center}
\includegraphics[width=0.4\textwidth]{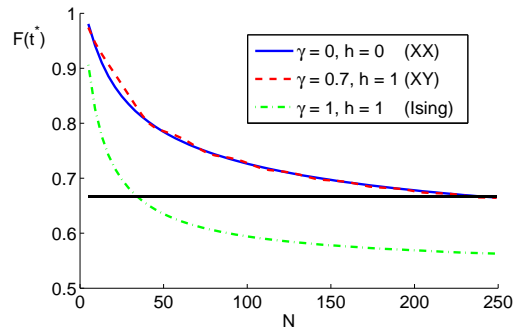}
\caption{(Color online) Scaling of the OAF at the arrival time $t^*$ versus length $N$, for different parameters of the Hamiltonian.}
\label{fig3}
\end{center}\end{figure}

The scaling of $ F^{\rm XY}(t^*)$ for increasing length $N$ is shown in
Fig.~\ref{fig3} where it is clear that for the best parameters
($\gamma=0$, $h=0$ and $\gamma=0.7$, $h=1$) the OAF decreases very slowly,
and it is greater than the classical value ($2/3$) even for chains up
to $N=240$. This can be proved in the XX case thanks to the analytical
results of appendix C. In fact, using Eq. \eqref{XX-bign}, we can show that
the solution of equation $F^{\rm XY}(t^*) = 2/3$ is $N=240$, in excellent
agreement with Fig.~\ref{fig3}.
The weak dependence of $ F^{\rm XY}(t^*)$ on $N$ for the best parameters
strengthens the statement that these indeed define the least
dispersive channel, no matter the length.
Conversely, for non optimal parameters (for example $\gamma=1$, $h=1$)
the OAF decreases quickly and becomes lower than the classical threshold value $2/3$ for chains longer than $N=32$.

Let us now consider the entanglement distribution.
We prepare
the qubit pair $0'$-$0$ in a pure entangled state, and let qubit
0 to be connected with the chain, while keeping $0'$ isolated,
so as to generate an entangled pair $0'$-$N$. We use
the concurrence \cite{concurrence} to quantify the amount of
entanglement in
$\rho_{_{0'N}}(t)$.

The density matrix of the qubit pair $0'$-$N$,
Eq.~\eqref{rho_0N}, is given by the Choi matrix as
$\rho_{_{0'N}}=\frac12 \mathcal{C}_\mathcal{E}$. The concurrence of
this state is \begin{equation}\label{concurrence} C(t)=\max\{0,\; \hat
C(u(t),w(t)), \; \hat C(w(t),u(t)) \} \end{equation} where,
\begin{equation} \label{concurrence2} \hat C(x,y) = x-\sqrt{(y^2+A(t))
(1-x^2-A(t))}. \end{equation} One can see that, at variance with the OAF,
entanglement distribution
depends on the initial state as it is a function of $A(t)$;
this is due to the fact that
during the dynamics the initial state of qubits 0 and $0'$
interferes with the initial state of the chain and deteriorates the
quality of the transmission.

In the following, we investigate different initial states of the channel
in order to
find out to which state a better quality transmission might correspond.
In particular, we will refer to two fully separable states, namely
the ferromagnetic state, with all the spins aligned along the
$z$ direction, e.g. $|0,0,...,0\ra$, and the N\`eel state,  with
neighbouring spins antiparallel to each other, e.g.
$|0,1,0,1,...,0,1\ra$.
Moreover, we will also study two different entangled
initial states, namely that defined by a series of singlet states, and
the ground state of the channel Hamiltonian.

Let us first consider the XX ($\gamma=0$) model, so as to exploit the
analytical expressions available (see appendix C).
We can prove that the concurrence achieves its maximum value, i.e.
$u(t)$, when $|\psi_{\rm ch}\ra$ is initialized in a ferromagnetic state.
In fact in this case, since $W_{ij}(t) \equiv 0$, it is
\begin{equation}\label{XXAt}
  A(t) =
\sum_{j,l=1}^N U_{Nj}(t)^*U_{Nl}(t)\la \psi_{\rm ch}|c_j^\dagger c_l |\psi_{\rm ch}\ra.
\end{equation}
which is equal to 0 (1) when $|\psi_{\rm ch}\ra$ consists in a tensor product
of down (up) spins.
For other initial states, in which $0<A(t)<1$, the concurrence is lower than $u(t)$.

\begin{figure}
\centering
    \includegraphics[width=.5\textwidth,angle=0]{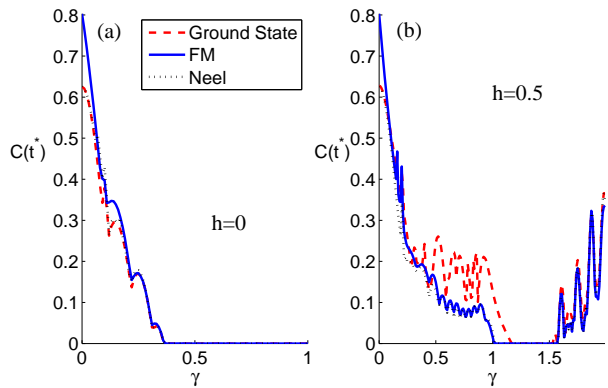}
    \caption{(Color online) Entanglement versus $\gamma$ in a chain of length $N=20$ for different initial states while the magnetic field takes: (a) $h=0$ and (b) $h=0.5J$.}
     \label{fig4}
\end{figure}

In Fig.~\ref{fig4}(a) we plot the concurrence as a function of anisotropy
$\gamma$ when $h=0$ for different initial states. As the figure clearly
shows increasing the anisotropy decreases the quality of transmission.

In the limit of $\gamma \rightarrow 1$ the Hamiltonian $H_{\rm ch}$
becomes Ising-like which has a poor transmitting quality. As one can see,
in
Fig.~\ref{fig4}(a), in the absence of magnetic field the
ferromagnetic initial state always gives the highest entanglement; in
particular, when anisotropy $\gamma$ is small the difference between this
initialization and the others is evident.

One may improve the poor
ability of entanglement distribution in highly anisotropic chains
(large $\gamma$) by switching on the magnetic field.
This is clearly seen in Fig.~\ref{fig4}(b) where we set $h=0.5$.
Furthermore, we notice that the field does not essentially
affect the transmission for small $\gamma$  and, quite surprisingly,
makes the ground state the best possible initial state for strongly
anisotropic chains.

\begin{figure}
\centering
    \includegraphics[width=.5\textwidth,angle=0]{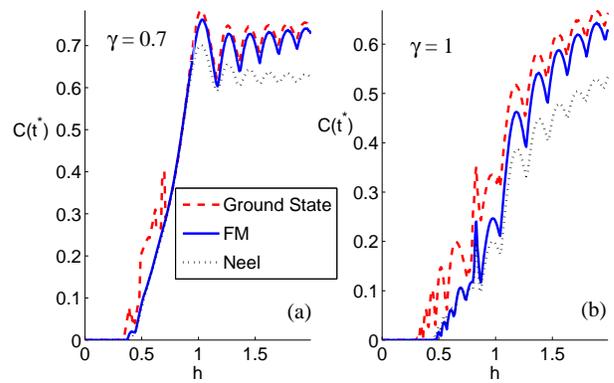}
    \caption{(Color online)
    Entanglement versus $h$ in a chain of length $N=20$ for different initial states while the anisotropy parameter takes: (a) $\gamma=0.7$ and (b) $\gamma=1$.}
     \label{fig5}
\end{figure}

In order to better understand
the role of magnetic field  we consider the transmitted entanglement
for different initial states as a function of $h$.
In Fig.~\ref{fig5} we plot the concurrence  for $\gamma=0.7$ and $\gamma=1$
and see that there exists a value of the field, slightly depending on
the initial state of the chain, above which the transmission becomes possible,
even for large anisotropies.

\begin{figure}\begin{center}
\includegraphics[width=0.5\textwidth]{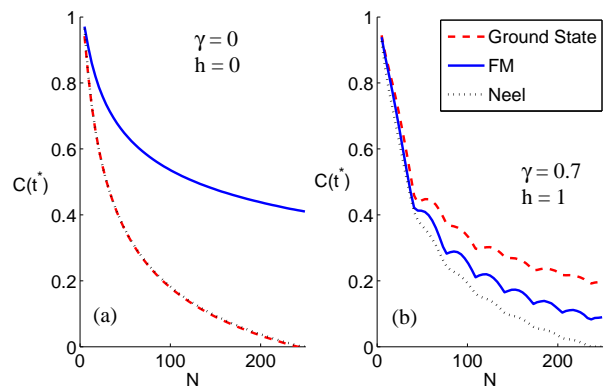}
\caption{(Color online) Scaling of the obtained entanglement at the arrival time versus length $N$ for different initial states: (a) isotropic XX Hamiltonian ($\gamma=0$ and $h=0$); (b) anisotropic XY Hamiltonian ($\gamma=0.7$ and $h=1$).}
	\label{fig6}
\end{center}\end{figure}

The existence of an exact solution for the XY Hamiltonian allows us
to study the quality of transmission for very long chains. In Fig.~\ref{fig6} we plot $C(t^*)$ as a function of $N$ for different initial states,
which evidently differentiate the entanglement transmission through long chains.
In particular, in Fig.~\ref{fig6}(a), we see that in the XX chain
the ferromagnetic initial state not only gives the highest concurrence
amongst the different initializations but it also provides the best scaling
with $N$.
In Fig.~\ref{fig6}(b) we plot $C(t^*)$ as a function of $N$ for $\gamma=0.7$
and $h=1$, i.e. for the parameteres that defines the less dispersive XY-like
channel (see Fig.~\ref{fig2}).
At variance with the state transmission case, where for such parameters
the transmission quality is as high as in the XX case
(see Fig.~\ref{fig3}),
during entanglement distribution through an XY chain we can not avoid
the interference $A(t)$ by properly choosing the initial state of the chain,
and a strong dependence on the length $N$ appears;
 Fig.~\ref{fig6}(b) in fact shows that this gives rise to
a significant lowering of the transferred concurrence.

Results for the N\`eel state and the series of singlets are found to be
very close to each other (therefore only those for the former state are
plotted): This shows that the inherent entanglement in the initial state
has a very little effect when a state is attached at one end of a spin chain.

\section{Interacting Systems: XXZ Hamiltonian} \label{XXZ}

After studying the effect of the initial state on the transmission
quality of the channel in free fermionic systems, interacting models
must also be considered, not only because they represent the
large majority of many-body systems, but also because they are usually
characterized by an extremely rich, though often difficult to be
physically deciphered, phenomenology.

Amongst the interacting Hamiltonians
the XXZ spin model here discussed, and defined by Eq.~(\ref{Hch}) with
\begin{equation}
  J_x = J_y = \frac{J}{2}, \h J_z = \frac{J\Delta}{2}~,
\end{equation}
is known to describe  many real systems and compounds,
thus playing an essential role, especially in one dimensional physics
\cite{mikeska,Lukin2003}.

This model has a very rich phase diagram: $\Delta<-1$ is the
ferromagnetic phase with a simple separable ground state with all
spins aligned to the same direction. For $-1<\Delta\leq1$ the
Hamiltonian
is gapless and the region is called $XY$ phase. $\Delta>1$ defines the
{\em N\`eel
phase}, where the spectrum is gapped and a finite staggered
magnetization arises. In the Ising limit $\Delta\gg 1$ the ground state is
the N\`eel state.

\begin{figure}
\centering
\includegraphics[width=0.5\textwidth]{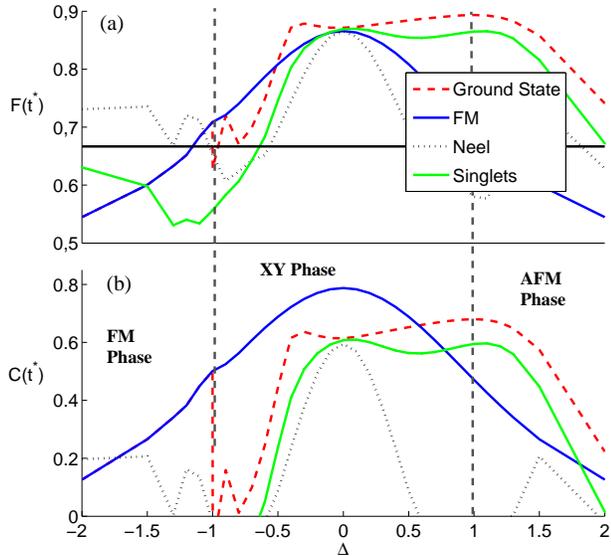}
    \caption{(Color online)  Transferring properties of a XXZ chain with the length of $N=20$ in its whole phase diagram for different initial states: (a) optimized average fidelity at the arrival time $t^*$
(the black line indicates the classically obtained fidelity $2/3$);
(b) obtainable entanglement at the arrival time $t^*$.}
     \label{fig7}
\end{figure}

Dynamical transmission in interacting fermionic models is very
different
from the free fermionic case, as it results from a complex combination
of many different effects amongst which the scattering
between interacting excitations and the existence of localized states.
As a matter of fact, there is no transmission process which
is generally independent on the initial state of the channel.
In particular we find that, depending on the value of $\Delta$, the best
transmission processes correspond to different initializations of the chain.

Let us first discuss the state transfer process.
In Fig.~\ref{fig7}(a) we see that even the OAF strongly depends on
$\ket{\psi_{\rm ch}}$: for $\Delta > 0$ the ground state clearly
gives the best possible initialization, while for $\Delta<0$ the
phenomenology is much more complicated.
When considering the entanglement distribution in Fig.~\ref{fig7}(b)
the situation is somehow
reversed, with a specific state, namely the ferromagnetic one,
granting the best possible initialization for $\Delta<0$, and a more
complex scenario for $\Delta>0$.

The general
phase diagram in Fig.~\ref{fig7} embodies the complex balance
between the effect of interference, which can be varied by acting on the
initial state of the chain, and the dispersiveness of the channel which
depends on $\Delta$.
As we already noted, in the XXZ case also the OAF is affected by
interference, and the overlap of $F(t^*)$ observed in Fig.~\ref{fig7}(a)
in the $\Delta=0$ point is due to its corresponding to a non-interacting model,
for which, as proved in section \ref{sec_XY}, the OAF does not depend on
the initial state.
On the other hand, there is no $\Delta$ value where the entanglement
distribution is independent on the destructive effects of interference.
Therefore, the only mechanism for removing such effect is that of choosing
a ferromagnetic initial state, whose dynamics can
be resolved in the single particle sector \cite{bose} where
 interference does not occur.  The absence of interference
is the reason why for $\Delta=0$ the $C(t^*)$ obtained with the
ferromagnetic initial state is by far larger then the others.
Finally, in Figs.~\ref{fig7}(a) and (b) we see that, given the initial state,
$F(t^*)-2/3$ and $C(t^*)$ have a similar behaviour as a function of $\Delta$,
with a shift in $C(t^*)$ for the FM initial state, consistent with the
exact analysis given above for the XX case.

Indications about the effects of dispersiveness of the channel on the
transmission processes can also be extracted: the ground state is seen
to minimize such effects for whatever $\Delta$, though this implies a better
entanglement distribution only for $\Delta>0.5$ where interference probably
plays a minor role leaving dispersion as the main destructive effect.

\section{Conclusion} \label{conclusion}
In this paper we have studied the quality of state and entanglement
transmission through quantum channels described by spin chains
varying both the Hamiltonian parameters and the initial state of the channel.
We have considered a vast class of Hamiltonians, including interacting
and non-interacting fermionic systems, which contains some of the most
relevant experimental realizations of one dimensional many-body systems,
both in the framework of solid state physics and in the realm of cold atoms
in optical lattices.

We find that if a free-fermionic model is available and an XY-like spin
Hamiltonian can be effectively realized, then the best possible tuning
of the parameters is that corresponding to the XX model with a ferromagnetic
initial state, both for state and entanglement transfer, whose quality
stays surprisingly high even for chains as long as $N \simeq 240$.
In the anisotropic case, a state transfer of the same quality as that attained
at $\gamma=0$, is obtained for $\gamma=0.7$ and $h=1$: referring to the framework developed in Ref. \cite{BACVV2010}, we infer that the relevant excitations lie in the linear zone of the dispersion relation and the resulting dynamics is essentially dispersionless.
Moreover,
good results for both state and entanglement transmission are
found in a wide range of the parameters $\gamma$ and $h$, providing
the channel is initialized in its ground state.
When an interacting XXZ model with a specific $\Delta$ is at hand,
one has to choose whether to
optimize the state or the entanglement transmission, since these goals
are obtained with different initial states.
In fact, we find that
the optimal average fidelity is more sensitive to the dispersiveness of the
channel, while the entanglement distribution is more sensitive to the
interference with the initial state of the chain.
As a matter of fact
the former gets its maximum in the antiferromagnetic isotropic
($\Delta=1$) channel initialized in its ground state, while the latter is
maximized by an XX ($\Delta=0$) channel initialized in a ferromagnetic state.

Our analysis show that the fidelity and the
entanglement do not necessarily quantify
the quality of quantum communication in the same way.
Namely, highest entanglement transfer can occur along a spin chain
which is different from that giving the highest average fidelity.
In fact, to the best of our knowledge, these results are the first
example
in which state and entanglement transmission show different features
due to the different role played in
such processes by dispersion, essentially set by the parameters of the
Hamiltonian, and interference, which explicitly depends on
the initial state of the channel. However, when we have higher entanglement one can always purify/distill entanglement using local operations \cite{bennett-distillation} and subsequently use it for teleportation and eventually end up with a higher fidelity.

{\em Acknowledgement:-} LB and PV gratefully acknowledge usefull
discussions with T.~J.~G.~Apollaro, A.~Cuccoli, and R.~Vaia. AB and SB
acknowledge the EPSRC. SB also thanks the Royal Society and the Wolfson
Foundation.

\appendix

\section{Optimal average fidelity}
In this section we derive an explicit expression for calculating the
optimal average fidelity (OAV) defined in Eq.~\eqref{Fav1}. Using the
results of Ref.~\cite{horofidel}
\begin{align}\label{appa1}
\frac32 & \max_{R\in U(2)} \int d\psi_{_0}
\bra{\psi_{_0}}R^\dagger\mathcal{E}\left[\ket{\psi_{_0}}\bra{\psi_{_0}}\right]R\ket{\psi_{_0}} -\frac12 = \nonumber\\ &
\max_{R\in U(2)} \bra{\Phi^+} (I\otimes R^\dagger\,\mathcal{E})
\left[\ket{\Phi^+}\bra{\Phi^+}\right] (I\otimes R)\ket{\Phi^+} = \nonumber\\ &
\max_{\ket{\psi_{\rm me}}} \;
\bra{\psi_{\rm me}} (I\otimes\mathcal{E})
\left[\ket{\Phi^+}\bra{\Phi^+}\right] \ket{\psi_{\rm me}},
\end{align}
since $(I\otimes R) \ket{\Phi^+}$, for varying unitary matrix $R$,
generates every maximally entangled state $\ket{\psi_{\rm me}}$.
Moreover, in \cite{badziag} it was proved that
\begin{equation}\label{appa2}
\max_{\ket{\psi_{\rm me}}} \bra{\psi_{\rm me}} \rho \ket{\psi_{\rm me}} =
\frac14 \left(1+t_1+t_2-\sign(\det(T))t_3\right)
\end{equation}
where $T_{mn}=\Tr[\rho\,\sigma_m\otimes\sigma_n]$ and $t_i$ are the
singular values of $T$, i.e. the square root of the eigenvalues of
$T^T T$, in decreasing order.
The explicit expression of equations \eqref{Fav2} and \eqref{mmat}
follows directly from Eqs.~\eqref{appa1},\eqref{appa2}
and for the properties of Choi matrix
$\mathcal{C}_\mathcal{E}\equiv (I\otimes\mathcal{E})
\left[2 \ket{\Phi^+}\bra{\Phi^+}\right]$
which can be easily proved from its definition \eqref{choi}.
In fact,
\begin{align}
\mathcal{M}^\mathcal{E}_{mn} & =
\Tr[\sigma_m \mathcal{E}(\sigma_n)] =
\Tr\left[\mathcal{C^{\mathcal{E}}} \sigma_n^T \otimes \sigma_m\right] =
\nonumber\\& =
(-1)^{n+1}\Tr\left[\mathcal{C^{\mathcal{E}}} \sigma_n \otimes \sigma_m\right].
\end{align}

\section{Diagonalization of the XY chain}
The total XY Hamiltonian is
\begin{equation}\label{H_XY}
    H_{\gamma} = J\sum_{k=0}^{N-1} \{ \sigma_k^+ \sigma_{k+1}^- 
    +\gamma\sigma_k^+ \sigma_{k+1}^+ 
    +\text{h.c.} \} + h\sum_{k=0}^N \sigma_k^z.
\end{equation}
which, by substituting the spin operators with their fermionic counterparts
takes the form
\begin{equation}\label{H_fermion}
  H=\sum_{k,l=0}^N \{ c_k^\dagger A_{kl} c_l +\frac{1}{2} (c_k^\dagger B_{kl} c_l^\dagger- c_k B_{kl} c_l)  \},
\end{equation}
where, $A$ is a symmetric matrix with elements $A_{kl}=J(\delta_{k,l+1}+\delta_{k,l-1})+h\delta_{k,l}$ and $B$ is an anti-symmetric matrix with elements $B_{kl}=\gamma(\delta_{k,l-1}-\delta_{k,l+1})$.
The above  quadratic Hamiltonian can be diagonalized
in the form \eqref{Q_diag} using a Bogoliubov transformation \eqref{eta},
and the diagonalization process reduces to finding the energies $E_k$ and the
matrices $P$ and $Q$.
From Eq.~\eqref{Q_diag}
\begin{equation}\label{commutation}
[H,d_k]=E_k d_k^\dagger.
\end{equation}
and then, by using Eq.~\eqref{H_fermion} and \eqref{eta} the conditions read
\begin{eqnarray}\label{A-B-h-g}
AP^T+BQ^T&=&+P^TE \cr
AQ^T+BP^T&=&-Q^TE
\end{eqnarray}
where, $E$ is a diagonal matrix with elements $E_{kl}=E_k \delta_{k,l}$.
To ensure that the transformation \eqref{eta} be canonical and invertible
the matrices have to satisfy these conditions
\begin{eqnarray}\label{g-h}
PP^T+QQ^T&=&P^TP+Q^TQ=\bar{I}_{N+1} \cr
PQ^T+QP^T&=&P^TQ+Q^TP=0,
\end{eqnarray}
which can be simplified by defining the new matrices
$\alpha=P+Q$ and $\beta=P-Q$. In fact, Eqs.~\eqref{g-h} force $\alpha$
and $\beta$ to be orthogonal matrices.
Moreover, since $(A+B)^T = A-B$, Eqs.~\eqref{A-B-h-g} simplifies into
a single equation
\begin{equation}\label{diagsvd}
A-B=\alpha^TE\beta
\end{equation}
which is the singular value decomposition of the matrix $A-B$.
The desired matrices $P=(\alpha+\beta)/2$ and $Q=(\alpha-\beta)/2$ are computed accordingly and the diagonalization process is completed.

%
%
%
%

For completeness we show also the inverse transformation that easily comes
from \eqref{g-h}
\begin{eqnarray}\label{inverse-eta}
c_k&=&\sum_{l=0}^N P_{_{lk}} d_l + Q_{_{lk}} d_l^\dagger, \cr
c_k^\dagger&=&\sum_{l=0}^N P_{_{lk}} d_l^\dagger + Q_{_{lk}} d_l.
\end{eqnarray}
%

\section{Analytical evaluation of $u(t)$ in the XX model}
In the XX model we have $E_k = 2J\cos(\frac{\pi(k+1)}{N+2})+2h$,
$P_{kn} = \sqrt{\frac{2}{N+2}} \sin(\frac{\pi(k+1)(n+1)}{N+2})$
and $Q_{kn} = 0$, making $W_{kn}(t) \equiv 0$.
Since the magnetic field causes only a constant shift in the dispersion
relation, in the following we set $h=0$. In this case
\begin{align}\label{XX-parameters}
  U_{_{N0}}(t) &= \sum_{k=0}^N e^{-itE_k} P_{k0} P_{kN}
  \nonumber\\
	  & = \sum_{m=1}^\infty i^{-m} \mathcal{J}_m(\beta)
	  \sum_{k=0}^{N} \cos\left(\frac{\pi m (k+1)}{N+2}\right) P_{k0}P_{kN}
  \nonumber\\
	  &\simeq i^{-N} \big(\mathcal{J}_N(\beta)+2\mathcal{J}_{N+2}(\beta)+\mathcal{J}_{N+4}(\beta)\big)
  \nonumber\\
	  &= \frac{4}{i^{N}}\left[\left(\frac{N+2}{\beta}\right)^2 \mathcal{J}_{N+2}(\beta)-
	  \frac{\mathcal{J}_{N+2}'(\beta)}{\beta}\right]
\end{align}
where $\beta=2Jt$. In the above equation  we used the Jacobi-Anger expansion and some properties of the Bessel function $\mathcal{J}_{n}$  \cite{abramowitz}.
The approximation consists in neglecting the Bessel
functions of order $m (N+2)$, with $m\ge 2$, since they contribute only after times of order $\frac{mN}{2J}$.
In fact, one can show that at the transmission time $t^*$,
i.e. the time when $U_{_{N0}}(t)$ takes its first peak,
$\beta^* \equiv 2Jt^* \simeq N -\xi (N/2)^{\frac13}$. Using the properties
of Bessel function \cite{abramowitz}
\begin{equation}\label{XX-asymp}
u(t^*) = \frac{2^{\frac73}}{N^{\frac13}}\airy(\xi) +
\frac{2\xi}{5N}\left(3\xi\airy'(\xi)+22\airy(\xi)\right)
\end{equation}
where $\airy(\xi)$ is the Airy function.
It can be proved \cite{bose} that the maximum of $u(t^*)$ is reached
for $\xi = -1.019$ and thus
\begin{equation}\label{XX-bign}
u(t^*) = \frac{2.700}{N^{\frac13}}-\frac{4.804}N.
\end{equation}

\end{document}